\pdfoutput=1
\documentclass{article}

\PassOptionsToPackage{numbers}{natbib}

\usepackage[preprint]{amlc_guide}

\usepackage[utf8]{inputenc} % allow utf-8 input
\usepackage[T1]{fontenc}    % use 8-bit T1 fonts
\usepackage{lmodern}        % Type 1 outline fonts (avoids Type 3 bitmaps)
\usepackage{hyperref}       % hyperlinks
\usepackage{url}            % simple URL typesetting
\usepackage{booktabs}       % professional-quality tables
\usepackage{amsfonts}       % blackboard math symbols
\usepackage{nicefrac}       % compact symbols for 1/2, etc.
\usepackage{microtype}      % microtypography
\usepackage{xcolor}         % colors
\usepackage{amsmath}
\usepackage{graphicx}
\usepackage{tabularx}
\usepackage{longtable}
\usepackage{CJKutf8}        % Japanese character support
\usepackage{multirow}
\usepackage{algorithm}
\usepackage{algorithmic}

\title{Low-Latency Spell Correction for Japanese Music Search Queries}

\author{
  Anshul Garg \and Pavni Tandon \and Karan Bhukar \and
  Tanmay Khandelwal \and Ujjal Kumar Dutta \\
  Amazon \\
  \texttt{\{gaanshu, paavnee, bhukar, tanmayx, ujjalkrd\}@amazon.com}
}

\begin{document}

\maketitle

\begin{abstract}
Spell correction for Japanese search queries presents unique challenges due to the
co-existence of four writing scripts (Latin/romaji, hiragana, katakana, and kanji) and
the distinct error patterns each script induces. We present a compact BART-based
sequence-to-sequence model (3 encoder + 3 decoder layers) designed for low-latency
spell correction of Japanese music search queries. The core contribution lies in a
script-aware synthetic misspelling generation pipeline that produces realistic training
data by combining keyboard-layout models (QWERTY and flick input), phonetic confusion
priors mined from real query logs, voiced/unvoiced consonant alternations, and
kana case errors. A key design decision is normalizing mixed-script catalog titles
to a single canonical script before misspelling synthesis, which we show is critical
for reducing model hallucinations. We train a custom byte-level BPE tokenizer on the
target music catalog to handle all four scripts in a unified vocabulary. Experiments
on a curated evaluation set show that our model achieves an exact-match accuracy
of 41.09\% and a character error rate (CER) of 11.62\%, outperforming edit-distance baselines
and achieving the lowest character error rate among all evaluated systems while maintaining sub-4ms
inference latency on a single GPU. We further analyze performance across individual
scripts and mixed-script queries, demonstrating the effectiveness of script-aware
data augmentation through systematic ablation studies.
\end{abstract}

\section{Introduction}

Search query spell correction is a critical component of modern information retrieval
systems, directly impacting user experience and downstream retrieval quality. While
spell correction for English and other Latin-script languages has been extensively
studied~\cite{norvig2007spell, symspell2012}, Japanese poses a fundamentally different
challenge: users may type queries in any of four writing scripts (hiragana, katakana,
kanji, and Latin/romaji) or in arbitrary combinations thereof. Each script introduces
its own characteristic error patterns. For instance, hiragana and katakana input via
flick keyboards on mobile devices produces adjacency errors that differ entirely from
QWERTY typos in romaji, while kanji input errors often stem from homophone confusion
during input method editor (IME) conversion.

In the domain of music search, these challenges are amplified. Music entity names
frequently contain mixed-script text (e.g., an artist name in katakana followed by a
song title in kanji with romaji loanwords), and users often search using a different
script than the canonical catalog entry. A robust spell correction system must therefore
handle all four scripts and their combinations within a single model while meeting
strict latency requirements for real-time search.

We address this problem with three contributions:

\begin{enumerate}
    \item \textbf{Script-aware synthetic data generation.} We design a misspelling
    synthesis pipeline that models error distributions specific to each Japanese script,
    including flick keyboard adjacency errors for kana, QWERTY typos for romaji,
    voiced/unvoiced consonant confusion (\begin{CJK}{UTF8}{min}が\end{CJK} $\leftrightarrow$
    \begin{CJK}{UTF8}{min}か\end{CJK}), small/large kana alternation
    (\begin{CJK}{UTF8}{min}っ\end{CJK} $\leftrightarrow$
    \begin{CJK}{UTF8}{min}つ\end{CJK}), and contextual phonetic priors mined from
    real query-entity correction logs.

    \item \textbf{Script normalization for training data.} We show that normalizing
    mixed-script catalog titles to a single canonical script before generating
    synthetic misspellings is critical for reducing model hallucinations on
    mixed-script inputs. Our search index stores entity titles in pure script
    (single writing system per title), and aligning the training data generation
    with this constraint significantly improves correction quality.

    \item \textbf{A compact seq2seq architecture.} We train a 6-layer BART model
    (3 encoder + 3 decoder layers) with a custom byte-level BPE tokenizer built on
    the music catalog vocabulary, enabling unified handling of all scripts with
    sub-4ms inference latency. We demonstrate that data quality compensates for
    model size, achieving near-LLM correction quality at two orders of magnitude
    lower latency.
\end{enumerate}

\section{Related Work}

\subsection{Spell Correction Approaches}

Traditional spell correction methods rely on edit distance computations against a
known dictionary~\cite{damerau1964technique, levenshtein1966binary}. Systems such as
SymSpell~\cite{symspell2012} achieve high throughput by pre-computing delete-based
neighborhoods, but their effectiveness degrades for languages like Japanese where
character-level edit distance poorly captures the phonetic and script-conversion
errors that dominate real user mistakes.

Neural approaches to spell correction have gained traction in recent years.
Encoder-decoder models based on the Transformer architecture~\cite{vaswani2017attention}
have been applied to grammatical error correction (GEC) in English~\cite{rothe2021simple}
and Chinese~\cite{zhang2020spelling}. The BART model~\cite{lewis2020bart}, pre-trained
as a denoising autoencoder, is particularly well-suited for spell correction as its
pre-training objective directly mirrors the correction task. Recent work has explored
adapter-based fine-tuning of BART for single-word spell correction in English, but
Japanese multi-script correction remains underexplored.

\subsection{Japanese-Specific Challenges}

Japanese text processing introduces unique difficulties stemming from its multi-script
writing system. Hagiwara and Mita~\cite{hagiwara2020github} highlighted the challenges
of Japanese GEC, noting that error patterns differ substantially across scripts.
Kato et al.~\cite{kato2023lattice} proposed the Lattice Path Edit Distance, a
romanization-aware edit distance metric for extracting misspelling-correction pairs
from Japanese search query logs, demonstrating that standard edit distance is
insufficient for Japanese. Their work at EMNLP 2023 showed that accounting for the
phonetic relationships between scripts significantly improves misspelling pair extraction.

For Japanese search queries specifically, prior work has explored query alteration
based on semantic similarity~\cite{suzuki2009japanese} and statistical machine
translation approaches. However, these methods typically handle only a single script
or require separate models per script, increasing system complexity and latency.

\subsection{Synthetic Data for Spell Correction}

The generation of synthetic misspellings for training data has been explored in
several contexts. Belinkov and Bisk~\cite{belinkov2018synthetic} studied synthetic
noise for machine translation robustness. For spell correction specifically,
approaches range from random character perturbations to keyboard-layout-aware noise
injection~\cite{pruthi2019combating}. Our work extends these ideas to the Japanese
multi-script setting, incorporating script-specific error models grounded in real
user behavior patterns observed in production query logs.

\section{Methodology}

\subsection{Problem Formulation}

We formulate Japanese spell correction as a sequence-to-sequence task. Given a
potentially misspelled query $q = (c_1, c_2, \ldots, c_n)$ where each character $c_i$ may
belong to any of the four Japanese scripts or punctuation, the model produces a
corrected sequence $\hat{q} = (\hat{c}_1, \hat{c}_2, \ldots, \hat{c}_m)$. When the
input query is already correctly spelled, the model should produce an identity mapping
($\hat{q} = q$).

\subsection{Script-Aware Synthetic Data Generation}
\label{sec:data_generation}

The core novelty of our approach lies in the synthetic misspelling generation pipeline.
We start from correctly spelled music entity titles from the catalog and generate
realistic misspellings using a combination of script-specific techniques.

\subsubsection{Script Identification and Conversion}

Each input title is first classified into one of five categories: \textsc{Latin},
\textsc{Hiragana}, \textsc{Katakana}, \textsc{Kanji}, or \textsc{Mixed}. For
mixed-script titles, we normalize to a single target script following a fixed priority
order: katakana $>$ hiragana $>$ Latin $>$ kanji. The target script is selected as
the highest-priority script present in the input. For example,
\begin{CJK}{UTF8}{min}``トうキょウ''\end{CJK} (katakana + hiragana) normalizes to
\begin{CJK}{UTF8}{min}``トウキョウ''\end{CJK} (pure katakana). This design reflects
the product requirement that spell correction can be offered in any script as long as
that script is present in the input query. Transliteration is performed using the
\texttt{pykakasi}~\cite{pykakasi} and \texttt{jaconv}~\cite{jaconv} libraries.

\subsubsection{Error Generation Techniques}

For each title, we apply 1 to $\min(5, \lceil 0.3 \times |q| \rceil)$ errors
sampled from the following technique pool, with uniform weights across all applicable
techniques:

\paragraph{Keyboard Layout Errors.}
For Latin-script text, we model QWERTY keyboard adjacency errors by replacing a
character with a randomly selected neighbor on the QWERTY layout. For kana text,
we model flick keyboard errors using a flick input adjacency map that captures the
spatial relationships of the Japanese flick keyboard commonly used on mobile devices.

\paragraph{Phonetic Confusion Priors.}
We mine contextual phonetic confusion rules from two sources: (1) real query-to-entity
(Q2E) correction logs - these are the records of which catalog entity a user ultimately engaged
with for a given search query from the music search system, and (2) the Japanese Wikipedia
Typo Dataset (JWTD)~\cite{jwtd2020}. For each observed correction pair
$(q_{\text{misspelled}}, q_{\text{correct}})$, we extract the minimal edit with its
surrounding context in the format \texttt{left\_ctx[from>to]right\_ctx}. For example,
the rule \begin{CJK}{UTF8}{min}\texttt{ん[ほ>ぽ]\$}\end{CJK} indicates that
\begin{CJK}{UTF8}{min}ほ\end{CJK} is replaced by
\begin{CJK}{UTF8}{min}ぽ\end{CJK} (handakuten addition) with probability 0.3\%
when preceded by \begin{CJK}{UTF8}{min}ん\end{CJK} at the end of the query.
These contextual rules are aggregated and weighted by their empirical probability.
At generation time, applicable rules are sampled proportionally to their observed
frequency. We maintain separate prior distributions for Latin (romanized) text and
for each Japanese script (hiragana, katakana, kanji), with the top 1000 rules per
category.

\paragraph{Voiced/Unvoiced Consonant Alternation.}
Japanese kana characters have systematic voiced (\begin{CJK}{UTF8}{min}濁点\end{CJK},
dakuten) and semi-voiced (\begin{CJK}{UTF8}{min}半濁点\end{CJK}, handakuten) variants.
We model errors where users omit or incorrectly add voicing marks, e.g.,
\begin{CJK}{UTF8}{min}か\end{CJK} $\leftrightarrow$ \begin{CJK}{UTF8}{min}が\end{CJK}
or \begin{CJK}{UTF8}{min}は\end{CJK} $\leftrightarrow$ \begin{CJK}{UTF8}{min}ば\end{CJK}
$\leftrightarrow$ \begin{CJK}{UTF8}{min}ぱ\end{CJK}.

\paragraph{Small/Large Kana Alternation.}
Japanese uses small kana variants (\begin{CJK}{UTF8}{min}っ, ゃ, ゅ, ょ\end{CJK})
that are visually similar to their full-size counterparts
(\begin{CJK}{UTF8}{min}つ, や, ゆ, よ\end{CJK}). We model substitution errors
between these pairs.

\paragraph{Generic Perturbations.}
We additionally apply script-agnostic perturbations: character deletion, character
repetition, adjacent character transposition, and space removal. These capture
universal typing errors that occur regardless of script.

\subsubsection{Data Composition}

We source correctly spelled titles from the music catalog: top 100K artists, albums,
podcasts, and playlists, top 500K tracks, and 1M most clicked entities from the Q2E
index (ranked by 28 day engagement), yielding 2.8M unique titles from 1.2M entities.
For each entity, we collect titles in multiple locale variants when available.

For each title, we generate 50-100 synthetic misspellings, filtering out those that
coincide with valid catalog entries. We additionally incorporate $\sim$3M real
misspelling--correction pairs from Q2E logs, filtered to pairs with edit distance
below $\min(5, 0.35 \times |q|)$ on romanized forms. The training set includes
identity pairs at a 1:4 ratio (misspelled to correct) to teach the model to preserve
already-correct queries.

\subsubsection{Script Normalization}
\label{sec:script_norm}

A key design decision is normalizing mixed-script catalog titles to a single
canonical script before generating synthetic misspellings. Since our search index
stores entity titles in pure script, the model's task is always to map a user query
to a pure-script catalog form. We normalize using the priority order described in
Section~\ref{sec:data_generation}, then generate misspellings from these pure-script
forms, so the model learns a consistent mapping: \emph{pure-script misspelling
$\rightarrow$ pure-script correction}. Without this step, mixed-script training
inputs force the model to simultaneously correct errors and normalize scripts,
leading to increased hallucination (Section~\ref{sec:ablation}).

\subsection{Tokenizer}

We train a custom byte-level BPE tokenizer~\cite{sennrich2016neural} on the music
catalog titles with a vocabulary size of 50{,}000 tokens. Byte-level BPE naturally
handles all four Japanese scripts plus Latin characters, numbers, and special
characters without requiring script-specific preprocessing. The tokenizer is trained
with special tokens \texttt{<s>}, \texttt{<pad>}, \texttt{</s>}, \texttt{<unk>},
and \texttt{<mask>}, and uses a maximum sequence length of 20 tokens, which covers
over 99.9\% of music search queries.

\subsection{Model Architecture and Training}

We use a BART~\cite{lewis2020bart} encoder-decoder architecture trained from scratch
with a custom byte-level BPE tokenizer, as our domain-specific vocabulary differs
substantially from standard pre-trained BART models. The compact 6-layer architecture
is chosen to meet the sub-4ms latency requirement. Since the model is autoregressive,
inference latency scales with output length; the average output on our evaluation set
is 6.4 tokens, which mirrors the average token length in production search queries.

\begin{table}[h]
\centering
\caption{Model and training configuration.}
\label{tab:config}
\begin{tabular}{ll}
\toprule
\textbf{Parameter} & \textbf{Value} \\
\midrule
Encoder / Decoder layers & 3 / 3 \\
Hidden dim / Attention heads & 768 / 12 \\
FFN dimension & 3072 \\
Vocabulary size & 50{,}000 \\
Dropout / Activation & 0.1 / GELU \\
Beam size (inference) & 4 \\
\midrule
Max training steps & 30{,}000 \\
Effective batch size & 32{,}768 \\
Learning rate & $5 \times 10^{-5}$ \\
LR scheduler & Constant with warmup (0.05) \\
Precision / Optimizer & BF16 / AdamW \\
\bottomrule
\end{tabular}
\end{table}

\section{Experiments}

\subsection{Evaluation Data}

We construct an evaluation set of 1{,}618 real world Japanese music search queries
with zero overlap with the training data, covering all four scripts. The evaluation
set is stratified across script types: approximately 30\% Latin, 25\% hiragana,
25\% katakana, 10\% kanji, and 10\% mixed-script queries. It combines synthetically
generated misspellings with real misspelling-correction pairs mined from
query-entity (Q2E) search logs using romanization-aware edit distance filtering,
ensuring coverage of both systematic and naturalistic error patterns.
28.7\% of the evaluation queries
are already correctly spelled, testing the model's ability to avoid over-correction.

\subsection{Evaluation Metrics}

We evaluate using: \textbf{Exact Match (EM)}, the fraction of queries where the
model output exactly matches the gold correction; \textbf{Character Error Rate (CER)},
the character-level edit distance normalized by gold length
($\text{CER} = \text{EditDist}(\hat{q}, q^*) / |q^*|$);
\textbf{Correction F1}, the harmonic mean of correction precision (accuracy among
queries where the model changed its output) and correction recall (fraction of
misspelled queries correctly fixed); and \textbf{Pass-Through Accuracy (PT)}, the
fraction of already correct queries preserved unchanged.

\subsection{Baselines and Ablation Variants}

We compare against three external baselines: (1)~\textbf{SymSpell}~\cite{symspell2012}
applied to romanized queries and catalog entries with max edit distance~2;
(2)~\textbf{Lattice Path Edit Distance}~\cite{kato2023lattice}, a
romanization-aware method that finds the nearest catalog entry via lattice search;
and (3)~\textbf{LLM Zero-Shot (Claude Haiku)}, zero-shot prompting with catalog
context (see Appendix~\ref{app:llm_prompt}), serving as a quality upper bound at $\sim$800ms latency.

\label{sec:ablation}
To isolate each pipeline component, we train the same 6-layer BART with identical
hyperparameters on different training data: \textbf{Random noise only} (script-agnostic
perturbations, no keyboard or phonetic models); \textbf{QWERTY keyboard only}
(QWERTY adjacency errors plus generic perturbations, no Japanese-specific models);
and \textbf{No script normalization} (full error techniques but without mixed-to-pure
script normalization).

\subsection{Results}

\begin{table*}[t]
\centering
\caption{Spell correction performance on the Japanese music query evaluation set (n=1{,}618).
The top section shows external baselines; the bottom section shows ablation variants
of our system (same 6L BART architecture, different training data). Attempt\% is the
fraction of misspelled queries where the model changed its output. Best results among
latency-viable systems ($<$100ms) are in \textbf{bold}.}
\label{tab:main_results}
\begin{tabular}{lcccccccc}
\toprule
\textbf{System} & \textbf{EM (\%)} & \textbf{CER (\%)} & \textbf{F1 (\%)} & \textbf{PT (\%)} & \textbf{Attempt (\%)} & \textbf{Latency (ms)} \\
\midrule
\multicolumn{7}{l}{\emph{External baselines}} \\
\midrule
SymSpell (romanized)              & 22.19 & 25.95 & 8.69 & 65.09 & 84.23 & $<$1 \\
Lattice Path Edit Distance        & 20.46 & 31.65 & 6.54 & 59.27 & 29.46 & $\sim$500 \\
LLM Zero-Shot (Claude Haiku)      & 48.27 & 27.24 & 42.46 & 53.45 & 98.79 & $\sim$800 \\
\midrule
\multicolumn{7}{l}{\emph{Ablation variants (same 6L BART, different training data)}} \\
\midrule
Ours: Random noise only           & 25.73 & 17.22 & 9.61 & 74.57 & 72.76 & $<$4 \\
Ours: QWERTY keyboard only        & 18.73 & 42.42 & 4.48 & 74.14 & 80.55 & $<$4 \\
Ours: No script normalization     & 34.08 & 15.75 & 23.78 & 82.33 & 74.32 & $<$4 \\
\textbf{Ours: Full pipeline}      & \textbf{41.09} & \textbf{11.62} & \textbf{31.68} & \textbf{88.36} & 71.98 & $<$\textbf{4} \\
\bottomrule
\end{tabular}
\end{table*}

Table~\ref{tab:main_results} presents the overall results. We introduce a new metric,
Attempt\%, which measures the fraction of misspelled queries where the model changed
its output (i.e., attempted a correction). This metric reveals a fundamental difference
in system behavior: the LLM baseline attempts correction on 98.79\% of misspelled
queries, while our full pipeline attempts correction on only 71.98\%. Our model
achieves the lowest CER (11.62\%) among all systems while maintaining sub-4ms
inference latency, representing a $>$200$\times$ speedup over the LLM baseline.

\paragraph{External baselines.}
The SymSpell baseline achieves 22.19\% exact match with an aggressive
correction strategy (84.23\% attempt rate), but suffers from high CER (25.95\%) and
low pass-through accuracy (65.09\%), indicating frequent over-correction of
already-correct queries. The Lattice Path Edit Distance approach performs poorly
overall (EM 20.46\%, CER 31.65\%), as brute-force romanization matching scales
poorly with catalog size. The LLM baseline achieves the highest correction F1
(42.46\%) with near-universal attempt rate (98.79\%), but at $\sim$800ms latency
and with only 53.45\% pass-through accuracy, making it unsuitable for production.

A notable pattern is that the LLM achieves higher EM (48.27\%) than our model
(41.09\%) yet substantially worse CER (27.24\% vs.\ 11.62\%). The LLM gets more
queries exactly right, but when it errs, it errs catastrophically - it
over interprets short queries by expanding them into what it believes the user
meant, e.g., \begin{CJK}{UTF8}{min}ミスチル\end{CJK} $\rightarrow$ Mr.Children
(275\% CER), \begin{CJK}{UTF8}{min}ガガ\end{CJK} $\rightarrow$
\begin{CJK}{UTF8}{min}レディー・ガガ\end{CJK} (250\% CER),
\begin{CJK}{UTF8}{min}リサ\end{CJK} $\rightarrow$ LiSA (200\% CER). These are
semantically reasonable but produce outputs much longer than the gold, inflating CER.
Our model avoids this failure mode because it is conservative: when unsure, it
returns the input unchanged.

\paragraph{Ablation analysis.}
The ablation variants reveal the contribution of each pipeline component:
\begin{itemize}
    \item \textbf{Random noise only} vs.\ \textbf{Full pipeline}: The random noise
    variant achieves correction F1 of 9.61\% and CER of 17.22\%, confirming that
    linguistically-motivated, script-aware noise generation is essential.

    \item \textbf{QWERTY keyboard only} vs.\ \textbf{Full pipeline}: QWERTY-only
    augmentation yields F1 of 4.48\%, the lowest among ablation variants, confirming
    that off-the-shelf Latin keyboard augmentation is insufficient for Japanese.

    \item \textbf{No script normalization} vs.\ \textbf{Full pipeline}: The
    no-normalization variant achieves F1 of 23.78\%, substantially below the full
    pipeline (31.68\%), confirming that script normalization is critical.
\end{itemize}

The full pipeline achieves a correction F1 of 31.68\%, substantially higher than
the best ablation variant (no script normalization, 23.78\%), demonstrating the synergistic effect of combining all
script-aware techniques with script normalization.

\subsubsection{Per-Script Analysis}

\begin{table}[h]
\centering
\caption{Per-script breakdown for our full pipeline model.}
\label{tab:per_script}
\begin{tabular}{lccc}
\toprule
\textbf{Script} & \textbf{EM (\%)} & \textbf{CER (\%)} & \textbf{PT (\%)} \\
\midrule
Latin (romaji)  & 40.68 & 6.59 & 86.54 \\
Hiragana        & 43.64 & 15.76 & 88.70 \\
Katakana        & 46.15 & 9.30 & 97.06 \\
Kanji           & 51.11 & 12.10 & 97.37 \\
Mixed           & 29.34 & 13.29 & 69.01 \\
\midrule
Overall         & 41.09 & 11.62 & 88.36 \\
\bottomrule
\end{tabular}
\end{table}

Table~\ref{tab:per_script} shows the per-script breakdown. Kanji queries achieve
the highest exact-match accuracy (51.11\%) with excellent pass-through (97.37\%),
likely because kanji entities are highly distinctive in the catalog. Katakana queries
show strong performance (EM 46.15\%, CER 9.30\%) with the second-highest pass-through
rate (97.06\%), validating the effectiveness of our flick keyboard and dakuten error
models. Hiragana queries achieve EM of 43.64\% with higher CER (15.76\%), reflecting
the greater ambiguity among phonetically similar hiragana strings. Latin queries
achieve the lowest CER (6.59\%), reflecting the regularity of QWERTY typo patterns.
Mixed-script queries show the lowest EM (29.34\%)
and pass-through (69.01\%), as these require the model to handle multiple scripts
simultaneously.

\section{Discussion}

\paragraph{Data Quality Over Model Size.}
Our compact 6-layer model achieves the lowest CER (11.62\%) among all systems,
outperforming dictionary-based approaches (SymSpell: 25.95\%, Lattice: 31.65\%)
and the LLM baseline (27.24\%) at sub-4ms latency. The ablation study confirms
data quality as the primary performance driver, with the full pipeline achieving
F1 of 31.68\% versus 23.78\% for the next best variant.

\paragraph{Correction Behavior.}
Our model attempts correction on 71.98\% of misspelled queries, compared to 98.79\%
for the LLM and 84.23\% for SymSpell, resulting in high pass-through accuracy
(88.36\%). This conservative behavior is deliberate for production: over-correction
is more harmful to user experience than under-correction. Script normalization
contributes significantly-without it, the model must simultaneously correct errors
and normalize scripts, reducing F1 from 31.68\% to 23.78\%.

\paragraph{Limitations.}
The model operates at the query level without broader search context. Kanji
correction remains challenging due to semantic IME conversion errors. Our synthetic
data assumes correctly spelled catalog titles, which may not hold for user-generated
content such as playlist names.

\paragraph{Production Deployment.}
The model is hosted on a GPU instance using TensorRT-LLM for optimized inference
with BF16 precision, exposed via a managed serving endpoint. A guardrail file
containing high-traffic queries and their verified corrections is applied at serving
time to prevent mis-correction of popular queries, providing an additional safety
mechanism beyond the model's own confidence.

\section{Conclusion}

We presented a low-latency spell correction system for Japanese music search queries
that handles all four writing scripts within a single compact seq2seq model. The key
insight is that script-aware synthetic training data generated through keyboard-layout
models, phonetic confusion priors, and script normalization can compensate for both
data scarcity and compact model size. Ablation studies confirm that each pipeline
component contributes meaningfully, with the full system achieving F1 of 31.68\%
and the lowest CER (11.62\%) among all systems at sub-4ms latency. Future work
includes improving kanji IME error correction, curriculum learning for higher recall,
and knowledge distillation from larger models.

\section*{Customer Impact Statement}
This work directly improved the search experience for Japanese speaking users of a
large scale music streaming service. By correcting misspelled queries in real time
across all four Japanese writing scripts, the system reduced failed searches and
helped users find their intended content more reliably. The sub-4ms inference latency
ensures that spell correction adds no perceptible delay to the search experience.
The conservative correction strategy (high pass-through accuracy) is designed to
minimize user frustration from over-correction, prioritizing precision over recall
to avoid altering correctly spelled queries. The spell correction system was launched in production in Nov 2025 and it led to substantial increase in seconds listened in Japanese market which translated to material increase in revenue for Amazon Music from JP alone.

\appendix

\section{Qualitative Examples}
\label{app:examples}

Table~\ref{tab:examples} shows representative correction examples across scripts.

\begin{table}[h]
\centering
\caption{Representative spell corrections across scripts.}
\label{tab:examples}
\begin{tabular}{llll}
\toprule
\textbf{Script} & \textbf{Input} & \textbf{Output} & \textbf{Error} \\
\midrule
Latin & tatlor swift & taylor swift & QWERTY \\
Latin & twicefancy & twice fancy & Space \\
Hira. & \begin{CJK}{UTF8}{min}あぷる\end{CJK} & \begin{CJK}{UTF8}{min}あっぷる\end{CJK} & Gemination \\
Kata. & \begin{CJK}{UTF8}{min}フエスティバル\end{CJK} & \begin{CJK}{UTF8}{min}フェスティバル\end{CJK} & Small kana \\
Hira. & \begin{CJK}{UTF8}{min}よるしか\end{CJK} & \begin{CJK}{UTF8}{min}よるしか\end{CJK} & Pass-through \\
Kata. & \begin{CJK}{UTF8}{min}ヨルカ\end{CJK} & \begin{CJK}{UTF8}{min}ヨルシカ\end{CJK} & Deletion \\
\bottomrule
\end{tabular}
\end{table}

\section{Japanese Phonetic Prior Examples}
\label{app:priors}

Table~\ref{tab:jp_priors} shows the top contextual phonetic priors mined from the
Q2E dataset for hiragana. Each rule specifies a context-dependent character
substitution with its empirical probability.

\begin{table}[h]
\centering
\caption{Top 10 Japanese hiragana phonetic priors from Q2E logs.}
\label{tab:jp_priors}
\begin{tabular}{llcl}
\toprule
\textbf{Prior} & \textbf{Replacement} & \textbf{Prob. (\%)} & \textbf{Context} \\
\midrule
\begin{CJK}{UTF8}{min}か\end{CJK} & \begin{CJK}{UTF8}{min}が\end{CJK} & 0.94 & \begin{CJK}{UTF8}{min}り[か>が]と\end{CJK} \\
\begin{CJK}{UTF8}{min}て\end{CJK} & \begin{CJK}{UTF8}{min}で\end{CJK} & 0.49 & \begin{CJK}{UTF8}{min}ん[て>で]\$\end{CJK} \\
\begin{CJK}{UTF8}{min}た\end{CJK} & \begin{CJK}{UTF8}{min}だ\end{CJK} & 0.48 & \begin{CJK}{UTF8}{min}も[た>だ]ち\end{CJK} \\
\begin{CJK}{UTF8}{min}ほ\end{CJK} & \begin{CJK}{UTF8}{min}ぼ\end{CJK} & 0.43 & \begin{CJK}{UTF8}{min}ん[ほ>ぼ]\$\end{CJK} \\
\begin{CJK}{UTF8}{min}と\end{CJK} & \begin{CJK}{UTF8}{min}ど\end{CJK} & 0.41 & \begin{CJK}{UTF8}{min}\^{}[と>ど]ん\end{CJK} \\
\begin{CJK}{UTF8}{min}こ\end{CJK} & \begin{CJK}{UTF8}{min}ご\end{CJK} & 0.41 & \begin{CJK}{UTF8}{min}\^{}[こ>ご]め\end{CJK} \\
\begin{CJK}{UTF8}{min}き\end{CJK} & \begin{CJK}{UTF8}{min}ぎ\end{CJK} & 0.34 & \begin{CJK}{UTF8}{min}\^{}[き>ぎ]ゅ\end{CJK} \\
\begin{CJK}{UTF8}{min}は\end{CJK} & \begin{CJK}{UTF8}{min}ば\end{CJK} & 0.33 & \begin{CJK}{UTF8}{min}そ[は>ば]に\end{CJK} \\
\begin{CJK}{UTF8}{min}ほ\end{CJK} & \begin{CJK}{UTF8}{min}ぼ\end{CJK} & 0.32 & \begin{CJK}{UTF8}{min}\^{}[ほ>ぼ]く\end{CJK} \\
\begin{CJK}{UTF8}{min}と\end{CJK} & \begin{CJK}{UTF8}{min}ど\end{CJK} & 0.31 & \begin{CJK}{UTF8}{min}\^{}[と>ど]う\end{CJK} \\
\bottomrule
\end{tabular}
\end{table}

The priors reveal that the most common Japanese misspelling pattern is the omission
of voiced consonant marks (dakuten), particularly in contexts where the voiced form
is phonetically expected (e.g., after the nasal
\begin{CJK}{UTF8}{min}ん\end{CJK}).

\section{Flick Keyboard Layout}
\label{app:flick}

The Japanese flick keyboard arranges kana characters in a grid where each base
character can be flicked in four directions (up, down, left, right) to produce
related characters from the same consonant row. For example, flicking from
\begin{CJK}{UTF8}{min}あ\end{CJK} produces
\begin{CJK}{UTF8}{min}い, う, え, お\end{CJK}. Our flick adjacency map models
errors where the user's flick gesture lands on an adjacent character, producing
a character from a neighboring consonant row rather than the intended vowel variant.

\section{LLM Zero-Shot Prompt}
\label{app:llm_prompt}

The following prompt template was used for the Claude Haiku baseline. For each query,
\texttt{\{query\}} is replaced with the input and \texttt{\{catalog\_entries\}} with
the top-100 catalog titles retrieved by fuzzy match.

\begin{verbatim}
You are a Japanese music search spell
correction system. Given a potentially
misspelled music search query, output ONLY
the corrected query. If the query is
already correct, output it unchanged. Do
not add any explanation, just the corrected
text.

Query: {query}
Corrected:
\end{verbatim}

\end{document}